\documentclass[12pt,a4paper]{revtex4-2}
\usepackage[utf8x]{inputenc}
\usepackage{ucs}
\usepackage[english]{babel}
\usepackage{amsmath}
\usepackage{amsfonts}
\usepackage{amssymb}
\usepackage{graphicx}
\usepackage[left=2cm,right=2cm,top=2cm,bottom=2cm]{geometry}

\usepackage{indentfirst}

\usepackage{float}
\usepackage[caption = false]{subfig}
\begin{document}
\title{Wigner function dynamics with boundaries expressed as convolution}

\author{S. S. Seidov}
\affiliation{Theoretical Physics and Quantum Technologies Department, NUST ``MISIS'', Moscow, Russia}

\begin{abstract}
In the present paper a method of finding the dynamics of the Wigner function of a particle in an infinite quantum well is developed. Starting with the problem of a reflection from an impenetrable wall, the obtained solution is then generalized to the case of a particle confined in an infinite well in arbitrary dimensions. It is known, that boundary value problems in the phase space formulation of the quantum mechanics are surprisingly tricky. The complications arise from nonlocality of the expression involved in calculation of the Wigner function. Several ways of treating such problems were proposed. They are rather complicated and even exotic, involving, for example, corrections to the kinetic energy proportional to the derivatives of the Dirac delta--function. The presented in the manuscript approach is simpler both from analytical point of view and regarding numerical calculation. The solution is brought to a form of convolution of the free particle solution with some function, defined by the shape of the well. This procedure requires calculation of an integral, which can be done by developed analytical and numerical methods. 
\end{abstract}

\maketitle

\section{Introduction}
Phase space quantization is an approach to quantum mechanics in which one uses real--valued functions instead of operators \cite{Wigner}. The functions are related to the operators via Weyl transform and are called Weyl symbols \cite{Weyl}. The state of the quantum system is described by the Wigner function, which is the Weyl symbol of the density matrix. The product of two operators is replaced by a Moyal product of their Weyl symbols and the commutator is replaced by the Moyal bracket \cite{Moyal}. The latter can be expanded in powers of Plank constant $\hbar$ and the leading term is the classical Poisson bracket of the two functions. This makes the phase space quantization especially useful when studying quantum--classical correspondence \cite{dias_features_2007}.  

The phase space quantization formalism has a somewhat unexpected drawback: finding dynamics of the system, subjected to boundary conditions, appears to be nontrivial. In particular, the method of solving the dynamical equation first and applying the boundary conditions later does not provide a correct result. Thus solution of the textbook problems of a particle in an infinite box and of reflection from an impenetrable wall becomes an involved process \cite{walton_wigner_2007, zachos_features,kryukov_infinite_2005, belchev_robin_2010, dias_wigner_2002, dias_boundaries_2021}. This is due to nonlocal character of the Wigner function. In literature several ways of dealing with this problem are proposed. They are successful when solving for stationary states, but, however, are complex and hard to implement from computational point of view when solving for time evolution of an arbitrary initial state. In this paper a simpler method of solving the dynamical equations in phase space quantisation approach for a system with boundary conditions is proposed. It boils down to a convolution of two functions, which is a well--defined operation and can be performed by developed numerical algorithms if the analytical calculation is not possible.

The manuscript starts with a brief introduction to the phase space quantization formalism. Then the dynamics of a free particle in absence of the boundary conditions is described. Next the problem of reflection from an impenetrable wall is considered with three solutions: a naive and incorrect one, a solution via representing the potential wall as a limit of a smooth exponential function and a solution with additional terms in the kinetic energy of the Hamiltonian. Finally the main topic of the article is presented: solution via convolution and its generalization to a potential well in arbitrary dimensions.

\section{Phase space quantization}
This section is devoted to a brief introduction of the phase space quantization method. The reader is advised to see refs. \cite{polkovnikov_phase_2010, zachos_deformation_2002, curtright_concise_2014, curtright_quantum_2012, curtright_concise_2014, case_wigner_2008} for an in-depth review of the subject. 

\subsection{Weyl symbols of quantum--mechanical operators}
In the phase space quantization formalism the quantum system is described not by quantum operators, but by real--valued functions of phase space variables $x$ and $p$ --- coordinate and momentum of the system. These functions, called Weyl symbols of the operators, are related to the latter by a Weyl transform.  In particular, for an operator $\hat A$ its Weyl symbol is defined as
\begin{equation}
A(x, p) = \frac{1}{2\pi} \int\limits_{-\infty}^\infty e^{i p y} \left\langle x - \frac{y}{2}\right |\hat A \left| x + \frac{y}{2} \right \rangle d y.
\end{equation} 
Here $|x_0 \rangle$ is a quantum oscillator coherent state at point $x = x_0$ and $p = 0$ in the phase space. The Weyl symbol of a density matrix $\hat \rho$ is the Wigner function:
\begin{equation}
W(x, p) = \frac{1}{2\pi} \int\limits_{-\infty}^\infty e^{i p y} \left \langle x - \frac{y}{2} \right|\hat \rho \left| x + \frac{y}{2} \right\rangle d y.
\end{equation} 
If the quantum state is pure, i.e. the density matrix can be written in form $\hat \rho = |\psi\rangle \langle \psi|$, the Wigner function becomes
\begin{equation}\label{eq:W_def}
W(x, p) = \frac{1}{2\pi} \int\limits_{-\infty}^\infty e^{i p y} \psi^*\left(x - \frac{y}{2} \right) \psi \left(x + \frac{y}{2} \right) dy,
\end{equation}
where $\psi(x)$ is the wave function of the system in the coordinate representation. The Wigner function defines the probability densities of finding the system with a coordinate $x$ or momentum $p$:
\begin{equation}
\begin{aligned}
&P(x) = \int\limits_{-\infty}^\infty W(x, p) dp
&P(p) = \int\limits_{-\infty}^\infty W(x, p) dx.
\end{aligned}
\end{equation}

Essentially the Weyl symbol of an operator $\hat A$ is the Fourier transform of the expression $\langle x - y/2|\hat A|x + y/2\rangle$. This fact will become crucial for our further analysis.

\subsection{Moyal bracket}
The difference between classical and quantum mechanics arises from the difference between multiplication of classical phase space functions and of Weyl symbols of quantum operators. The Weyl symbol of the product of two operators $\hat A$ and $\hat B$ is the Weyl product of their symbols $A \star B$. It can be calculated as
\begin{equation}
A(x, p) \star B(x, p) = A\left(x + \frac{i}{2} \overrightarrow\partial_p, p - \frac{i}{2}\overrightarrow\partial_x \right) B(x, p) = A(x, p) B\left(x - \frac{i}{2}\overleftarrow\partial_p, p + \frac{i}{2}\overleftarrow\partial_x\right).
\end{equation}
The arrow above the derivative indicates the direction it acts on. Accordingly the Weyl symbol of the commutator $[\hat A, \hat B]$ is the Moyal bracket
\begin{equation}
\{\!\{ A, B \}\!\} = A \star B - B \star A.
\end{equation}

The von Neumann equation, defining the time evolution of the density matrix, after the Weyl transform becomes
\begin{equation}\label{eq:dotW}
\dot W = i \{\!\{H, W\}\!\}.
\end{equation} 
Here $H$ is the Weyl symbol of the Hamiltonian and the obtained equation defines time evolution of the Wigner function. It will be the subject of our study. 

\section{Dynamics of a free particle}
The Weyl symbol of the Hamiltonian of a free particle is
\begin{equation}
H = \frac{p^2}{2m}.
\end{equation}
Substituting it in (\ref{eq:dotW}) we obtain the equation for the dynamics of the Wigner function of a free particle:
\begin{equation}\label{eq:dotW_free}
\dot W_0(x, p, t) = \frac{i}{2m}\left(p - \frac{i}{2}\partial_x\right)^2 W_0(x, p, t) - \frac{i}{2m}\left(p + \frac{i}{2}\partial_x\right)^2 W_0(x, p, t) = - \frac{p}{m} \partial_x W_0(x, p ,t).
\end{equation}
We can formally integrate it obtaining the solution in form
\begin{equation}\label{eq:Wt}
W_0(x, p, t) = \exp \left\{-\frac{p t}{m} \partial_x \right\} W_0(x, p, 0) = W_0\left(x - \frac{p t}{m}, p, 0\right).
\end{equation}
Here we used the fact that $e^{\alpha \partial_x}$ is an operator of displacement in $x$ by amount $\alpha$. However, how it will be seen further, this should be done with caution.

Let us now calculate $\dot W$ using its definition (\ref{eq:W_def}). Although seemingly redundant at the moment, this calculation will become illustrative later when considering the system with boundary conditions. So,
\begin{equation}\label{eq:dW_int_1}
\dot W_0(x, p, t) = \frac{1}{2\pi} \int\limits_{-\infty}^\infty e^{i p y}\left[\dot \psi^*\left(x - \frac{y}{2}\right) \psi\left(x + \frac{y}{2}\right) + \psi^*\left(x - \frac{y}{2} \right)\dot \psi\left(x + \frac{y}{2} \right)\right] dy.
\end{equation}  
Taking into account Schroedinger equation $\dot \psi = -(i/m) \psi''$ we arrive at
\begin{equation}\label{eq:dW_int_2}
\begin{aligned}
\dot W_0(x, p, t) &= \frac{i}{2\pi m} \int\limits_{-\infty}^\infty e^{i p y}\left[{\psi^*}''\left(x - \frac{y}{2} \right) \psi\left(x + \frac{y}{2} \right) - \psi^*\left(x - \frac{y}{2} \right)\psi''\left(x + \frac{y}{2} \right)\right] dy = \\
& = -\frac{i}{2\pi m} \int\limits_{-\infty}^\infty e^{i p y} \partial_{xy}^2 \left[\psi^*\left(x - \frac{y}{2} \right) \psi\left(x + \frac{y}{2} \right)\right] dy =\\
&= -\frac{p}{2\pi m}  \partial_x \int\limits_{-\infty}^\infty e^{i p y} \left[\psi^*\left(x - \frac{y}{2} \right) \psi\left(x + \frac{y}{2} \right)\right] dy.
\end{aligned}
\end{equation}  
In the last step we used the property of the Fourier transform that $\mathcal{F}_y[i \partial_y u ](p) = p \mathcal{F}_y[u](p)$ for function $u(y)$. Thus we obtain $-(p \partial_x W_0)/m$ in the r.h.s and recover equation (\ref{eq:dotW_free}).

\section{Reflection of an impenetrable wall}
Now consider the problem with an impenetrable  wall placed at the origin, i.e. with the potential
\begin{equation}\label{eq:U}
U(x) = \begin{cases}
0, &x > 0\\
\infty, & x \leqslant 0
\end{cases}.
\end{equation}
The system is confined to the interval $x \in [0, \infty)$ and the wave function is subjected to the boundary condition $\psi(x \leqslant 0) = 0$. In the usual wave function formalism of the quantum mechanics we would solve the Schroedinger equation for $x \in (-\infty, \infty)$ and after that impose the boundary conditions. The wave function would be written in form $\psi(x) = \varphi(x)\theta(x)$, where $\theta(x)$ is the Heaviside function and $\varphi(0) = 0$. Further it will be shown that the same approach of simply multiplying the free particle Wigner function by $\theta(x)$ leads to incorrect results which do not satisfy the boundary conditions. This is a well--known problem and several solutions were proposed. However they are computationally intensive and in the current manuscript a simpler alternative is presented. 

\subsection{Naive and incorrect solution}
First let us try to act in the same way as in case of wave function formalism, i.e. we write the Wigner function at $t = 0$ as
\begin{equation}
\begin{aligned}
&W(x, p, 0) = F(x, p) \theta(x)\\
&F(0, p) = 0.
\end{aligned}
\end{equation}
Now we use (\ref{eq:Wt}) and obtain the time dependent Wigner function
\begin{equation}
W(x, p, t) = F\left(x - \frac{pt}{m}, p \right) \theta\left(x - \frac{pt}{m}\right).
\end{equation}
Obviously the solution is incorrect because now $W(x, p, t) = 0$ for $x \leqslant pt/m$ and not for $x \leqslant 0$, as required by the boundary condition.

The reason for this is that we applied the free particle solution to the problem with a potential energy. Although one might be deceived by it being zero on the half--line $x > 0$, it is still a potential nevertheless and should be taken into account properly when solving the dynamics equation (\ref{eq:dotW}). In the wave functions formalism the method works due to locality of the Schroedinger equation: the time derivative $\dot \psi(x)$ in its l.h.s. equals to $i\hat H \psi(x)$ at the same point $x$. Thus the region of $x \leqslant 0$, where the potential is infinite, has no effect on the region $x > 0$, where the potential is zero. But the expression for the Wigner function contains non--local product $\psi^*(x - y/2)\psi(x + y/2)$. So even for $x > 0$ there is interference from the region $x \leqslant 0$ for specific values of $y$. If the wave function at the forbidden region is not zero, which is the case if we first solve for the free particle, said interference will introduce error to the solution.

From a mathematical point of view, the mistake is in careless treatment of the $\exp(-p t \partial_x/m)$ operator in the free particle solution (\ref{eq:Wt}) as a displacement operator. This is correct only if it acts on the functions from the Schwartz space. If the function $u(x)$ does not belong to the Schwartz space, in general 
\begin{equation}
\mathcal{F}[\partial_x^n u](s) \neq (i s)^n \mathcal{F} [u](s).
\end{equation}
And this equality is crucial when proving that $e^{\alpha \partial_x} u(x) = u(x + \alpha)$. The differential operators on the compact support and corresponding quantization constitute a deep research field \cite{clark_quantum_1980, vincenzo_2008, vincenzo_2016, al-hashimi_canonical_2021}. 

%\subsubsection{Displacement operator and non--hermiticity of momentum operator on $\mathbb{R}^+$}

\subsection{Solution via including the potential energy term to the dynamics equation}
In ref. \cite{kryukov_infinite_2005} the authors proposed to model the potential (\ref{eq:U}) as
\begin{equation}
U(x) = \lim_{\alpha \rightarrow \infty} e^{-\alpha x}.
\end{equation}
They were able to solve for the stationary states, defined by equation
\begin{equation}\label{eq:HW}
H \star W = E W,
\end{equation}
the analogue of the stationary Schroedinger equation. The disadvantage of this method is that it leads not only to complicated nonlinear differential equations, but also to difference equations. In particular, the potential energy contributes to the dynamics equation (\ref{eq:dotW}) as following:
\begin{equation}
\begin{aligned}
&\lim_{\alpha \rightarrow \infty} \{\!\{e^{-\alpha x}, W(x, p) \}\!\} = \lim_{\alpha \rightarrow \infty} \left[e^{-\alpha \left(x + \frac{i}{2} \partial_p\right)} - e^{-\alpha \left(x - \frac{i}{2} \partial_p\right)} \right] W(x, p) = \\
&= -2 i \lim_{\alpha \rightarrow \infty} e^{-\alpha x} \sin \left(\frac{\alpha \partial_p}{2} \right) W(x, p) = \lim_{\alpha \rightarrow \infty}e^{\alpha x} \left[W\left(x, p - \frac{i \alpha}{2}\right) - W\left(x, p + \frac{i \alpha}{2} \right) \right].
\end{aligned}
\end{equation}
The authors find a way to deal with such terms when solving equation (\ref{eq:HW}) and end up with a well defined differential equation in $\alpha \rightarrow \infty$ limit. However, when solving dynamics equation (\ref{eq:dotW}) instead of the stationary equation (\ref{eq:HW}) these terms present a severe difficulty as it is hard to interpret them (the momentum gets shifted by $\pm i \infty$) and to solve the resulting equation. 

\subsection{Solution via correction of the kinetic part of the Hamiltonian}
An alternative approach was taken in refs. \cite{dias_wigner_2002, dias_boundaries_2021}. It consist in repeating calculations (\ref{eq:dW_int_1}) and (\ref{eq:dW_int_2}) but on the restricted interval. In application to our case of potential (\ref{eq:U})
\begin{equation}\label{eq:W_2x}
\begin{aligned}
W(x, p) &= \frac{1}{2\pi} \int\limits_{-\infty}^{\infty} e^{i p y} \varphi^*\left(x - \frac{y}{2}\right)\theta\left(x - \frac{y}{2}\right)\varphi\left(x + \frac{y}{2} \right) \theta \left(x + \frac{y}{2} \right)dy = \\
& = \frac{1}{2\pi} \int\limits_{-2x}^{2x} e^{i p y} \varphi^*\left(x - \frac{y}{2} \right)\varphi\left(x + \frac{y}{2} \right)dy.
\end{aligned}
\end{equation}
Here the wave function is represented as $\psi(x) = \varphi(x)\theta(x)$. By differentiating with respect to time we obtain the expression, which differs from (\ref{eq:dW_int_1}) only by the limits of integration (again the Schroedinger equation is taken into account):  
\begin{equation}\label{eq:dW_limited}
\begin{aligned}
\dot W &= \frac{i}{2\pi m} \int\limits_{-2x}^{2x} e^{i p y}\left[{\varphi^*}''\left(x - \frac{y}{2} \right) \varphi\left(x + \frac{y}{2} \right) - \varphi^*\left(x - \frac{y}{2}\right)\varphi''\left(x + \frac{y}{2} \right) \right] dy  =\\
& = -\frac{i}{2\pi m}  \int\limits_{-2x}^{2x} e^{i p y} \partial_{xy}^2 \left[\varphi^*\left(x - \frac{y}{2}\right)\varphi\left(x + \frac{y}{2}\right) \right]dy.
\end{aligned}
\end{equation}
Let us now determine how does multiplication by $p$ affect the Wigner function. When the limits of integration were infinite it was equivalent to differentiation of the expression under the Fourier transform with respect to $y$. Now for finite limits of integration using integration by parts
\begin{equation}
\begin{aligned}
&-\frac{i p}{2 \pi m}\int\limits_{-2x}^{2x} e^{i p y} \partial_x \left[\varphi^*\left(x - \frac{y}{2} \right)\varphi\left(x + \frac{y}{2} \right) \right]dy =\\
&= -\frac{1}{2 \pi m} e^{i p y} \partial_x \left[\varphi^*\left(x - \frac{y}{2} \right)\varphi\left(x + \frac{y}{2} \right) \right]\Bigg|_{-2x}^{2x} +  \frac{1}{2 \pi m} \int\limits_{-2x}^{2x} e^{i p y} \partial_{xy}^2 \left[\varphi^*\left(x - \frac{y}{2} \right)\varphi\left(x + \frac{y}{2} \right) \right]dy.
\end{aligned}
\end{equation} 
The second term is $i \dot W$, see eq. (\ref{eq:dW_limited}). Using the property of the derivative of the delta--function $\int f(x) \delta'(x - a) dx = -f'(a)$, the first term can be written in form
\begin{equation}
\begin{aligned}
&-\frac{1}{4m}\int\limits_{-2x}^{2x} e^{i p y}\varphi^*\left(x - \frac{y}{2}\right)\varphi\left(x + \frac{y}{2}\right) \left[\delta'\left(x - \frac{y}{2} \right) + \delta'\left(x + \frac{y}{2} \right) \right] dy =\\
&= -\frac{1}{4m} \{\!\{ \delta'(x), W(x, p) \}\!\}.
\end{aligned}
\end{equation}
Finally the dynamics equation for the Wigner function is
\begin{equation}
\dot W(x, p, t) = -\frac{p}{m}\partial_x W(x, p, t) - \frac{1}{4m}\{\!\{ \delta'(x), W(x, p, t) \}\!\}.
\end{equation}
The additional term might be thought as a contribution of a potential $U \sim \delta'(x)$. But in fact it is a correction to the kinetic part of the Hamiltonian (a sign of which is dependence of $m$) due to confinement of the system on a half--line $x > 0$ \cite{dias_boundaries_2021}. I.e. the Weyl symbol of the Hamiltonian of a free particle in this case becomes
\begin{equation}
H_{\mathbb{R}^+} = \frac{p^2}{2m} - \frac{1}{4m} \delta'(x) .
\end{equation}

Similar to previous case this is not a convenient way of describing the dynamics. The authors, when performing numerical calculations, approximate the delta--function as $\lim\limits_{\epsilon \rightarrow 0} \exp\{-(x/\epsilon)^2\}/(\sqrt{\pi} \epsilon)$. On the one hand, the approximation introduces additional numerical error. And on the other, when attempting analytical calculations, the Moyal bracket of the Wigner function with an exponent introduces infinite shifts of the momentum as in previous section.

\subsection{Boundary conditions as convolution}
By writing down the wave--function in form $\psi(x) = \varphi(x) \theta(x)$ and introducing shorthand notation functions $f(x, y, t)$ and $g(x,y)$ the Wigner function of the system can be represented as:
\begin{equation}\label{eq:Wfg}
\begin{aligned}
&W(x, p, t) = \int\limits_{-\infty}^{\infty} e^{i p y} f(x,y,t) g(x,y) dy\\
&f(x, y, t) = \varphi^*\left(x - \frac{y}{2}, t \right) \varphi\left(x  + \frac{y}{2}, t \right)\\
&g(x, y) = \theta\left(x - \frac{y}{2}\right) \theta\left(x  + \frac{y}{2}\right).
\end{aligned}
\end{equation}
We can see the Wigner function as a Fourier transform with respect to $y$ of a product:
\begin{equation}
W(x, p, t) = \mathcal{F}_y[f g](x, p, t).
\end{equation} 
The arguments of functions $f$ and $g$ were suppressed for brevity. The function $g$ does not depend on time, so $\dot W = \mathcal{F}_y[\dot f g](x, p, t)$. And the Fourier transform of the product equals to the convolution of the Fourier transforms, so 
\begin{equation}
\dot W(x, p, t) = \mathcal{F}_y[\dot f](x, p, t) *_p \mathcal{F}_y[g](x, p).
\end{equation} 
Here $*_p$ is the convolution of two functions with respect to variable $p$ (not to be confused with Moyal product $\star$):
\begin{equation}
u(p) *_p v(p) = \int\limits_{-\infty}^{\infty} u(k) v(p - k) dk.
\end{equation}
The last step is realising that $\mathcal{F}_y[\dot f](x, p, t) = \dot W_0(x, p ,t)$ is the time derivative of the Wigner function of the system in absence of the boundary conditions. For it the simple equation (\ref{eq:dotW_free}) and its solution (\ref{eq:Wt}) hold. Thus the solution for the boundary value problem can be constructed by convolution of the free solution with some function $\mathcal{F}_y[g](x, p)$ which depends on the exact form of boundary conditions:
\begin{equation}\label{eq:W_convolve}
W(x, p, t) = W_0\left(x - \frac{p t}{m}, p, 0 \right) *_p \mathcal{F}_y[g](x, p).
\end{equation} 

\subsubsection{Connection to method of images and calculation of $\mathcal{F}_y[g]$}
The problem of a quantum particle bouncing of a wall has an elegant solution via method of images \cite{andrews_wave_1998}, which consists of representing the wave function as
\begin{equation}\label{eq:psi_odd}
\psi(x, t) = [\phi(x, t) - \phi(-x, t)]\theta(x).
\end{equation}
Here $\phi(x, t)$ is the wave function satisfying the free Schroedinger equation $\dot \phi = -(i/m) \phi''$. Subtraction of the "image" $\phi(-x, t)$ makes sure that the function $\psi(x, t)$ satisfies the boundary condition $\psi(0, t) = 0$ $\forall t$ and is continuous at $x = 0$. The whole difference should be multiplied by $\theta(x)$ in order to cut off the unphysical non--zero values of the wave function at $x < 0$. 

Similar structure of the solution is obtained when using the proposed solution method via convolution. We rewrite the function $g(x, y)$ as
\begin{equation}
g(x, y) = \theta(x) \operatorname{rect}\left(\frac{y}{4 x} \right),
\end{equation} 
where $\operatorname{rect}(y/4x)$ is the rectangular function:
\begin{equation}
\operatorname{rect}\left(\frac{y}{4 x} \right) = \theta\left(\frac{y}{4x} + \frac{1}{2} \right) - \theta\left(\frac{y}{4x} - \frac{1}{2} \right) = 
\begin{cases}
1, & |y| < 2 x\\
0, & |y| > 2 x
\end{cases}.
\end{equation}
The Fourier--transform of $g(x, y)$ can now be easily calculated:
\begin{equation}
\mathcal{F}_y[g] = \frac{\sin(2 p x) \theta(x)}{\pi p} = \frac{2 x}{\pi} \theta(x) \operatorname{sinc}(2 p x).
\end{equation}
Here $\operatorname{sinc}(x) = (\sin x)/x$.

According to (\ref{eq:W_convolve}) the Wigner function of the particle bouncing of a wall is given by convolution of the free particle Wigner function $W_0(x, p, t)$ with the expression above:
\begin{equation}\label{eq:W0_sinc}
W(x, p, t) = \frac{2 x}{\pi} \theta(x) \int\limits_{-\infty}^{\infty} W_0(x, p - k, t) \operatorname{sinc}(2 k x) dk.
\end{equation}
If we now demonstrate that the integral is the odd function of variable $x$, the solution will be shown to have structure analogous to (\ref{eq:psi_odd}), i.e. an odd function multiplied by $\theta(x)$. Given that the integration is taken over $k$, we can restrict our analysis to the analysis of the properties of the integrand. The factor $x \operatorname{sinc}(2 k x)$ is an odd function of $x$, thus the whole product is odd if and only if $W_0(x, p - k, t)$ is an even function of $x$. This is, of course, not true for any function $W_0$, but imposes a condition on the allowed wave functions, corresponding to $W_0$. Indeed, $W_0 \sim \int e^{ipy} \psi^*(x - y/2) \psi(x + y/2) dy$, thus the product $\psi^*(x - y/2) \psi(x + y/2)$ should be an even function of $x$. This is only possible if both factors are either simultaneously even or simultaneously odd. Given condition $\psi(0) = 0$ the first option requires $\psi(x) \equiv 0$ and the non--trivial wave functions should be odd, exactly as they are constructed using method of images.

\subsubsection{Free solution as a limit far from the wall}
It is expected that for from the wall placed at the origin, i.e. in limit of $x \rightarrow \infty$, the solution (\ref{eq:W0_sinc}) will tend to the free solution (\ref{eq:Wt}). This can be verified using the property of the $\operatorname{sinc}$--function
\begin{equation}
\lim_{a \rightarrow \infty} \frac{a}{\pi} \int\limits_{-\infty}^{\infty} \operatorname{sinc}(a x) u(x) dx = u(0).
\end{equation}
Then taking the limit one confirms that indeed far from the wall the solution tends to the free particle one:
\begin{equation}
\lim_{x \rightarrow \infty} \frac{2 x}{\pi} \theta(x) \int\limits_{-\infty}^{\infty} W_0(x, p - k, t) \operatorname{sinc}(2 k x) dk = \theta(x) W_0(x, p , t).
\end{equation} 

\section{Generalization of the method}
The proposed method can be applied to an arbitrary potential made of impenetrable walls by choosing the appropriate function $g$ corresponding to the boundary conditions. For example, if the particle is placed in an infinite quantum well and the wave function is required to be zero outside of the interval $x \in [a, b]$, the function $g$ becomes
\begin{equation}
g(x, y) = \left[\theta\left(x - a - \frac{y}{2} \right) - \theta\left(x - b - \frac{y}{2}\right) \right] \left[\theta\left(x - a + \frac{y}{2} \right) - \theta\left(x - b + \frac{y}{2} \right) \right].
\end{equation}
In this case the same relation with the method of images remains, but now infinite number of "images" of the Wigner function appear at both sides of the well.

\begin{figure}[h!!]
\center\includegraphics[width = 0.5\textwidth]{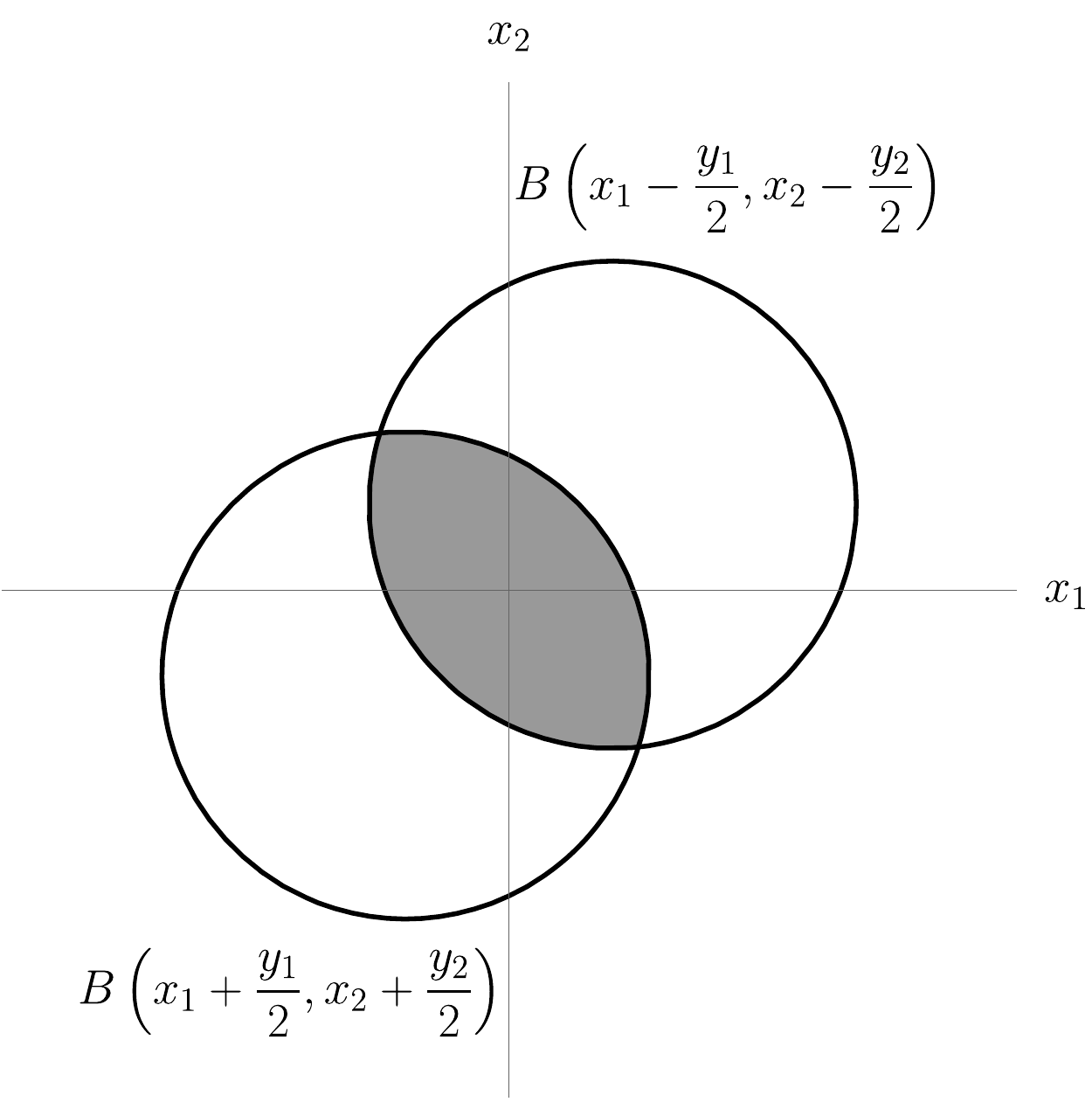}
\caption{Function $g(x_1, x_2, y_1, y_2)$ following from the boundary condition for a circular billiard: $g(x_1, x_2, y_1, y_2) = 1$ in the shaded region and $g(x_1, x_2, y_1, y_2) = 0$ elsewhere. Solid lines --- contours $B\left(x_1 \pm \dfrac{y_1}{2}, x_2 \pm \dfrac{y_2}{2} \right) = 1$.}
\label{fig:intersec}
\end{figure}

In higher dimensions more variety is possible --- the particle might be placed in quantum wells of different shapes. Suppose the boundary of the potential well is defined by an equation $B(x_1, x_2, \ldots, x_n) = 1$. For example, the equation $x_1^2 / R^2 + x_2^2 / R^2 = 1$ defines a two dimensional circular billiard with radius $R$. Then the function $g$ is 
\begin{equation}
\begin{aligned}
g(x_1, \ldots, x_n, y_1, \ldots, y_n) &= \theta\left(1 - B\left(x_1 - \frac{y_1}{2}, \ldots, x_n - \frac{y_n}{2} \right) \right)\times \\
&\times \theta\left(1 - B\left(x_1 + \frac{y_1}{2}, \ldots, x_n + \frac{y_n}{2} \right) \right).
\end{aligned}
\end{equation}
This function is unity in the intersection of regions $B(x_1 \pm y_1/2, \ldots, x_n \pm y_n/2) < 1$ and zero in the rest of the $\{x_1, \ldots, x_n\}$ space. The function $g$ corresponding to the two dimensional circular billiard is shown in fig. \ref{fig:intersec}.

Analytical calculation of the function $\mathcal{F}_y[g]$ for the potential wells of complicated form is usually not possible. However, the numerical procedure remains well defined and computationally efficient.

\section{Conclusions}
The nonlocality of the product $\psi^*(x - y/2) \psi(x + y/2)$, entering the expression for the Wigner function, does not allow to solve the boundary value problems buy just simply imposing boundary conditions on the general solution. Instead one should either "honestly" solve the problem considering the effect of the potential, or modify the equations of motion. The manuscript presents a method of construction of the boundary value problem solution by convolution of the solution of the free particle problem with some function. The latter is defined by the particular form of boundary condition.The method provides a procedure which does not involve complicated nonlinear differential and even difference equations or exotic additional terms to the kinetic energy. Instead in its core lays a simple convolution which is just an integral and can be calculated numerically or in some cases analytically. 

\section{Acknowledgement}
The work was supported by the federal academic leadership program ‘‘Priority 2030’’ (MISIS Strategic Project Quantum Internet).

%\bibliography{biblio}

\end{document}